\documentclass[unnumsec,webpdf,modern,large]{template-arXiv}%

\graphicspath{{Fig/}}
\usepackage{siunitx}

\usepackage{color}
\usepackage{hyperref, soul}
\definecolor{linkColor}{rgb}{1,0,0}
\definecolor{citeColor}{rgb}{1,0,0}
\hypersetup{pdfborder={0 0 0}, colorlinks=true,urlcolor=linkColor,citecolor=citeColor}

\usepackage[mathlines, switch]{lineno}

\begin{document}

\journaltitle{Microscopy and Microanalysis}
\DOI{DOI HERE}
\copyrightyear{2025}
\pubyear{2025}
\access{Advance Access Publication Date: Day Month Year}
\appnotes{Original Article}

\firstpage{1}


\title[]{Neutral but Impactful: Gallium Cluster-Induced Nanopores from Beam-Blanked Gallium Ion Sources}

\author[1,2,3,$\ast$]{Dana O. Byrne}
\author[3]{Stephanie M. Ribet}
\author[3]{Karen C. Bustillo}
\author[4]{Colin Ophus}
\author[2,3,5,$\ast$]{Frances I. Allen}

\authormark{Byrne et al.}

\address[1]{\orgdiv{Department of Chemistry}, \orgname{University of California, Berkeley}, \orgaddress{\postcode{94720}, \state{CA}, \country{USA}}}
\address[2]{\orgdiv{Department of Materials Science and Engineering}, \orgname{University of California, Berkeley}, \orgaddress{\postcode{94720}, \state{CA}, \country{USA}}}
\address[3]{\orgdiv{National Center for Electron Microscopy, Molecular Foundry}, \orgname{Lawrence Berkeley National Laboratory}, \orgaddress{\postcode{94720}, \state{CA}, \country{USA}}}
\address[4]{\orgdiv{Department of Materials Science and Engineering}, \orgname{Stanford University}, \orgaddress{\postcode{94305}, \state{CA}, \country{USA}}}
\address[5]{\orgdiv{California Institute for Quantitative Biosciences}, \orgname{University of California, Berkeley}, \orgaddress{\postcode{94720}, \state{CA}, \country{USA}}}

\corresp[$\ast$]{Corresponding authors. \href{email:email-id.com}{ dana$\_$byrne@berkeley.edu, francesallen@berkeley.edu}}


\abstract{Neutral atoms originating from liquid metal ion sources are an often-overlooked source of contamination and damage in focused ion beam microscopy. Beyond ions and single atoms, these sources also generate atom clusters. While most studies have investigated charged clusters, here we demonstrate that neutral clusters are also formed. These neutral clusters bypass the electrostatic beam blanking system, allowing them to impinge on samples even when the ion beam is blanked. We investigate this phenomenon using thin (\qty{\leq20}{nm}) freestanding membranes of hexagonal boron nitride, silicon, and silicon nitride as targets. Randomly dispersed nanopores that form upon neutral cluster exposure are revealed. The average nanopore diameter is \qty{\sim2}{nm} with a narrow size distribution, suggesting that the atom clusters have a preferred size. Various electron microscopy techniques are used to characterize the nanopores, including high-resolution transmission electron microscopy, multislice ptychography, and electron energy-loss spectroscopy. Finally, we show how electron irradiation in the transmission electron microscope can be used to both remove any amorphous material that may clog the pores and to controllably grow the pores to specific sizes. Tunable nanopores such as these are interesting for nanofluidic applications requiring size-selective membranes.}
\keywords{Liquid metal ion source, neutrals, atom clusters, nanopores}

\maketitle


\section{Introduction}

Focused ion beam (FIB) microscopes are used to image, modify and analyze materials on the micro- to nanoscale in diverse fields ranging from the semiconductor industry to biology~\citep{Hoflich2023}. For example, FIB milling is routinely employed to prepare thin specimens for transmission electron microscopy (TEM) and electron tomography~\citep{Mayer2007,Berger2023}, as well as to sequentially slice bulk samples for 3D volume imaging by scanning electron microscopy~\citep{Scheffer2020}. Using the FIB, ions can also be implanted site-selectively, for example, to fabricate qubits for solid-state quantum computing~\citep{Hollenbach2022}, and in combination with mass spectrometers, the FIB can be used to analyze material compositions by secondary ion mass spectrometry~\citep{Audinot2021}. In each application, the ability to focus, shape, and steer the beam using electromagnetic fields — enabled by the charged nature of the particles — is the key factor driving the technique's versatility and precision. 
However, FIB sources also produce neutral species, either neutral atoms or atom clusters. Depending on the application, these neutrals can be a nuisance, or more unusually, they can be leveraged. 

The liquid metal ion source (LMIS) is the most established FIB source and still the most widely used, albeit in recent years other source technologies have been gaining momentum that deliver a wider range of ion species, beam currents and energies~\citep{Hoflich2023}. In the LMIS, a liquid metal reservoir is used to coat a sharp tungsten needle, which is placed next to an extractor counter electrode. The needle is held at a positive potential relative to the grounded extractor, producing an electric field at the needle apex that in combination with surface tension effects forms the so-called Taylor cone. Field evaporation of ions from the tip of the Taylor cone occurs, producing a thin stream of ions that are accelerated to the extractor~\citep{Taylor1964DisintegrationField,Gomer1979AppliedSources}. Various (typically electrostatic) ion-optical elements focus, steer and shape the beam, with the beam current selected using a simple aperture. The metal in the reservoir can be any low melting point metal or metal alloy, but gallium is the most commonly used.

While FIB sources are designed to produce accelerated ions, a significant fraction of the emission consists of neutrals. For example, in early LMIS studies, it was measured that up to 50\% or more of the mass loss from the source could be attributed to neutrals, with a strong dependence on the operating current of the source~\citep{Mair1981MassSources,Mair1979Gallium-field-ionAnodes}. Reasons for the production of neutrals include non-ionizing evaporation of atoms from the source and charge exchange collisions between emitted ions and residual gas molecules in the column~\citep{Hoflich2023}.
Modern LMIS systems are designed to reduce these effects, but neutral emission cannot be eliminated entirely. For applications in which background exposure from neutrals cannot be tolerated at all, a chicane-type beam blanker system can be employed~\citep{Kagarice2017}.

The neutral species generated may consist of single atoms or clusters of atoms, often referred to as droplets, which are emitted directly from the Taylor cone. These clusters can be formed already as neutral entities or as weakly charged clusters that follow the conventional ion beam path with a narrow angular distribution~\citep{DCruz1985IonSources,Wagner1981DROPLETSOURCES.}. While some of the latter neutralize, others retain their charge.
It is generally thought that cluster emission is negligible, provided that the source operating current is low
(\SI{<10}{\micro A})~\citep{Culbertson1979ATOM-PROBESOURCE.,Wagner1981DROPLETSOURCES.}. Nevertheless, the sizes of the emitted clusters are not insignificant (up to tens of atoms)~\citep{Barr1987GalliumSource, Saito1989AtomsSource,Sakaguchi1991AluminumSource,Bhaskar1987EvidenceSource}, and if their kinetic energy is sufficient, they can be expected to cause localized structural damage at randomly distributed impact sites on the specimen.   

In this work, we investigate neutral clusters produced by a gallium LMIS FIB microscope, probing the interaction of these clusters with thin (\qty{\leq20}{nm}) specimens. In contrast to prior cluster emission studies, which tend to rely on mass spectrometry and thus focus on charged clusters, our approach exclusively probes neutral clusters. We achieve this using the conventional electrostatic beam blanker to prevent all charged species from reaching the specimen, allowing only the neutrals to pass. Using various target materials, we show that the neutral particles can create nanopores, the size of which (diameters of \SI{\sim2}{nm}) indicate that atom clusters as opposed to single atoms are responsible. 
Advanced electron microscopy characterization of the cluster-irradiated specimens is performed to detect implanted gallium and to examine the structure and morphology of the nanopores that are formed. Finally, we demonstrate how gallium cluster irradiation from the LMIS in combination with electron irradiation in the transmission electron microscope can be used to remove contamination and expand the pores to fabricate nanopores that may be used for future functional applications like molecular filtration. 

\section{Experimental Methods}

\subsection{Target materials}

All samples were freestanding membrane samples suitable for subsequent analysis by TEM. Multilayer 2D hexagonal nitride (hBN) flakes were obtained from bulk using the standard tape exfoliation method and transferred to polydimethylsiloxane (PDMS) gel. Contrast in optical microscopy with benchmarking by atomic force microscopy was used to identify flakes of thickness \qty{<15}{nm}. Suitable flakes were then transferred to holey silicon nitide (SiN$_x$) TEM grids using mechanical dry release PDMS gel transfer~\citep{Castellanos-Gomez2014}. Other materials used as targets were \qty{20}{nm} SiN$_x$ and \qty{5}{nm} silicon membranes, obtained commercially as TEM grids.  

\subsection{Irradiation with neutrals from Ga FIB source}

Samples were loaded into a Zeiss ORION NanoFab FIB microscope equipped with a Ga LMIS, centered, and tilted to \SI{54}{\degree} to align the sample normal with the Ga column axis. The column valve was opened, but the ion beam (acceleration potential \qty{30}{kV}) was left electrostatically blanked, meaning that only neutral species could reach the specimen. A relatively large current-selecting aperture giving \SIrange{14}{17}{nA} of (blanked) ion beam current was chosen. Then, without ever exposing the sample to ions, samples were exposed to neutral particles for durations ranging from 20 minutes to 20 hours. 

\subsection{Characterization by electron microscopy}

\subsubsection{HR-TEM}

High-resolution TEM (HR-TEM) of irradiated samples and non-irradiated controls was performed at \SI{80}{kV} in a double-aberration-corrected modified FEI Titan 80-300 microscope (the TEAM I instrument at the Molecular Foundry, LBNL) equipped with a Continuum K3 direct electron detector. Additional HR-TEM was carried out in an image-corrected FEI ThemIS equipped with a Ceta2 CMOS camera, also at \SI{80}{kV}. 

The diameters of individual impact sites in the irradiated samples (later characterized to be pores) were determined from line profile analysis of the HR-TEM images using ImageJ software. Histograms of the size distributions were plotted using \qty{0.2}{nm} bins and are also represented using Gaussian kernel density estimate smoothing. 

\subsubsection{STEM-EDS}

Scanning TEM energy-dispersive X-ray spectrometry (STEM-EDS) of irradiated samples was performed at \SI{300}{kV} in an FEI TitanX microscope equipped with a Bruker Super-X quadrature detector (solid angle \SI{0.7}{sr}). The convergence angle was \qty{10}{mrad}.
Spectra were acquired using a \qty{4}{nA} probe, \qty{9.2}{ms} total pixel dwell time, \qty{0.65}{nm} step size, \qty{165}{nm}$\times$\qty{165}{nm} scan size, for 10 minutes of scanning. Quantification analysis of implanted Ga was performed using Bruker ESPRIT software from both the K- and L-shell peaks after background subtraction. 

\subsubsection{Correlative ptychography and STEM-EELS}

The 3D structure of the nanopores was studied using multislice electron ptychography. From a single 4D-STEM scan (acquisition of one diffraction pattern with co-ordinates ($k_x,k_y$) for every pixel with co-ordinates ($x,y$) ~\citep{Ophus2019}), this allows one to recover 3D structural information with a depth resolution on the order of a few nanometers~\citep{chen2021electron, ribet2024uncovering}.
Data collection for the ptychographic reconstructions was performed on the TEAM I microscope operated at \SI{80}{kV}. 
The probe was defined by a~\qty{23}{mrad} convergence angle and data was collected on the Dectris Arina camera using a dose rate of \SI{\sim5e8}{e^-/nm^2/s} and a total dose of \SI{\sim5e7}{e^-/nm^2} (typical frame time \qty{0.1}{s}). 
The probe current was \qty{70}{pA}, scan step size \SI{0.46}{\AA}, pixel dwell time \qty{0.5}{ms}, and the probe was defocused by $\sim$\SI{20}{nm}. 

The reconstructions were computed using a gradient descent algorithm in the open-source toolkit \texttt{py4DSTEM}~\citep{varnavides2023iterative}.
Initial parameters and calibrations for the multislice reconstruction were solved using Bayesian optimization routines in \texttt{py4DSTEM}.
The final potential reconstruction was performed using 25 \SI{2}{nm} slices with 400 iterations with a 0.1 step size in batches of 256 probe positions at a time.
Position correction as well as total variation de-noising was applied along the $xy$ and $z$ directions, with decreasing $z$ regularization after each 100 iterations.
Boundary conditions were applied to enforce positive potentials such that the top and bottom slices were 0 potential before applying total variation denoising.
Further information about the reconstruction formalism can be found elsewhere~\citep{varnavides2023iterative}.

Correlative STEM electron energy-loss spectroscopy (STEM-EELS) of the same nanopores investigated by ptychography was performed by operating in probe-corrected, energy-filtered STEM mode with a \qty{23}{mrad} convergence angle at \SI{80}{kV} using the same electron dose rate. The probe size for the EELS measurements was \qty{\sim0.1}{nm}, and the scan step sizes for the spectrum image acquisitions ranged from \SI{0.05}{}-\SI{0.1}{nm}. The dwell time per pixel was \qty{10}{ms}. We focused on the low-loss region, from \qty{0}{}--\qty{60}{eV}. A typical spectrum image took \qty{100}{s} to acquire. The spectral resolution was \qty{\sim1}{eV}, since the monochromator had been off for the ptychography data collection in the same experiment. 

Data analysis of the spectrum images was performed in \texttt{HyperSpy}~\citep{HyperSpy} using spatial binning followed by spectral smoothing with a total variation smoothing algorithm (weighted value of 5). Principal component analysis was used to create spatial masks to extract the spectral signals corresponding to the interior and exterior pore regions. These spectra were normalized to the $\pi$ plasmon peak at $\sim$\qty{7.5}{eV} before plotting.

\subsection{Electron irradiation of nanopores for cleaning and expansion}

Removal of amorphous material present on many of the nanopores was achieved using the FEI ThemIS electron microscope operated in parallel beam mode at \qty{80}{kV}. To track progress, images were acquired approximately every \qty{150}{s}, condensing the beam between frames in order to increase the electron dose rate and hence accelerate the unclogging and pore expansion process. The electron dose rate used for cleaning and growth was estimated to be \SI{\sim4e6}{e^-/nm^2/s} and the accumulated dose between frames was \SI{\sim6e8}{e^-/nm^2}. We note that in the STEM-mode ptychography and EELS experiments described above, pore expansion and/or obvious removal of amorphous material was not observed.

\section{Results and discussion}

\begin{figure}[h]
\centering
\includegraphics[width = 0.47\textwidth]{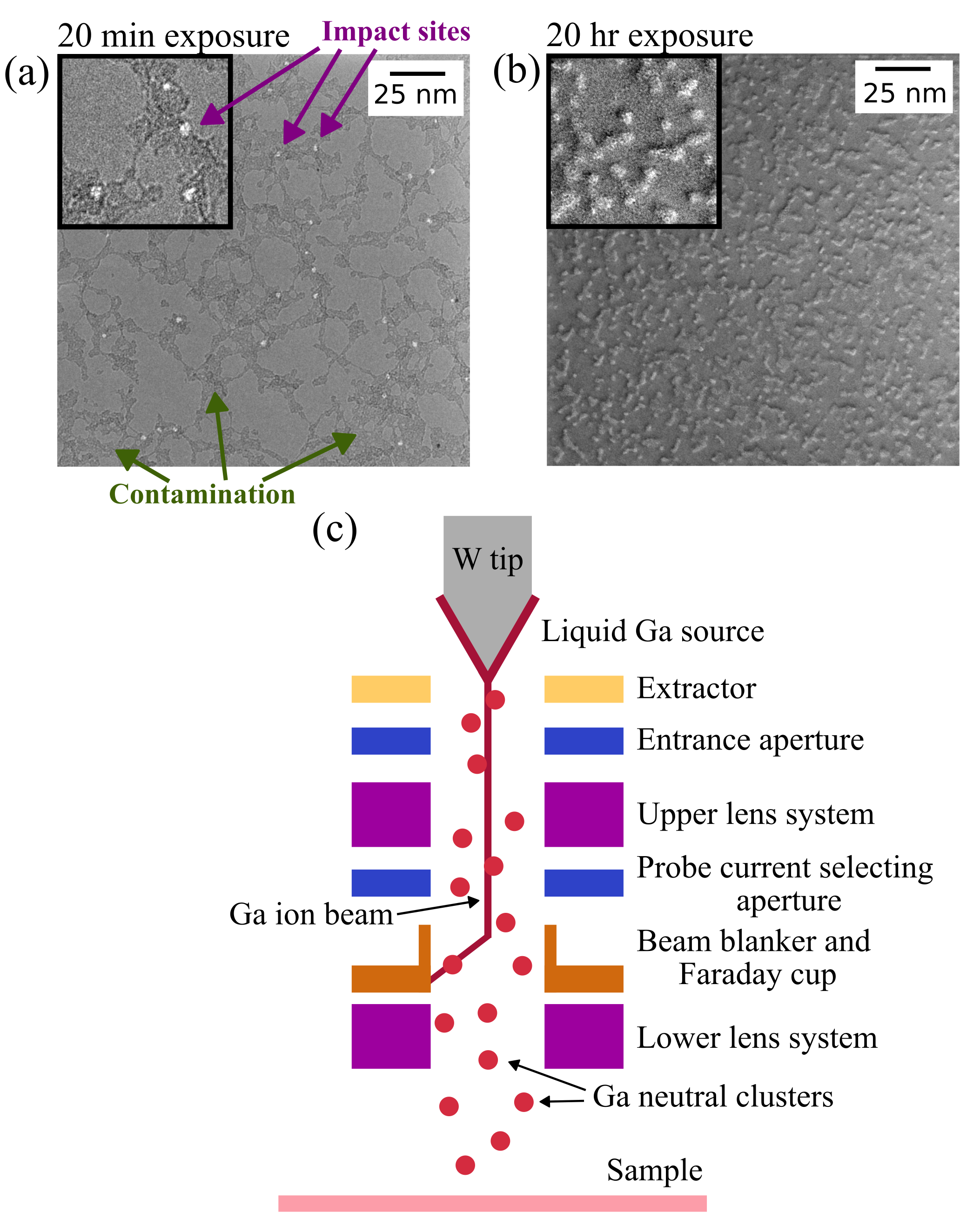}
  \caption{Bright-field TEM images of multilayer hBN samples showing the spatial distribution of pore-like structures obtained after (a) 20 minutes and (b) 20 hours of exposure to neutrals under the electrostatically blanked Ga LMIS. (c) Schematic illustrating the transit of Ga neutral clusters though the FIB column (here, assuming neutralization at the source), showing how the neutrals are unaffected by the blanker field and other electrostatic beam elements. The clusters then impact the specimen forming nanopores by local sputtering at the individual impact sites.}
  \label{fig:low mag}
\end{figure}

\subsection{Initial inspection of Ga neutral irradiated samples}

Figures ~\ref{fig:low mag}(a) and (b) show bright-field TEM images of two different hBN multilayer samples ($\sim$\SI{7}{}-\SI{15}{nm} thick) that had been placed under the Ga source with the column valve open and the ion beam electrostatically blanked for 20 minutes and 20 hours, respectively. After 20 minutes of exposure, a distribution of nanoscale brighter contrast regions, \qty{\sim1}{}-\qty{3}{nm} in diameter, is observed. Since these are bright-field images, we infer that these features likely correspond to regions of localized mass loss from the hBN. After 20 hours of exposure, a significantly higher density of these features is observed, many of which are spaced so closely that they form larger structures. 

A dark-field STEM image of the 20 minute exposed hBN sample is shown in Supplementary Fig.~S1. In addition, dark-field STEM images of similarly exposed \qty{5}{nm} Si and \qty{20}{nm} SiN$_x$ samples are shown in Supplementary Fig.~S2. All three sample types exhibited similar results, with each showing distributions of pore-like structures (presenting as dark contrast in the STEM images). Control samples that had not been exposed did not exhibit these features. 

The pore-like structures observed suggest localized damage from discrete particle impacts. Since the Ga column was open but the ion beam was electrostatically blanked, and given there was no other particle source active, we infer that the particles responsible are Ga neutrals generated at the Ga source and/or in the Ga column. Moreover, since the impact sites approach several nanometers in size, we conclude that the neutral particles are energetic atom clusters---in the case of the cluster, there is a high concentration of collisions and energy deposition near the surface, which can create significant peripheral damage~\citep{Aoki2010,Anders2005}, whereas in the case of a single impinging atom, the collision cascade path is very narrow, which would create much smaller (sub-nanometer) features~\citep{Thiruraman2018}. 

As mentioned in the Introduction, it is indeed known that in addition to ions, the Ga LMIS produces neutral particles (both single atoms and clusters). A schematic illustrating the fact that the flight path of such neutrals is unaffected by the electrostatic beam blanker in the FIB column is shown in Fig.~\ref{fig:low mag}(c). Whereas the Ga ions produced by the LMIS are deflected by the beam blanker, allowing the ion beam current to be measured using an off-axis Faraday cup, any neutrals transit the blanker field without deflection. Similarly, the electrostatic lenses, which focus, steer and shape the ion beam, have no effect on the neutrals. Thus, only the physical apertures will control the spread of the neutrals. 

\begin{figure*}
\centering
\includegraphics[width = 0.85\textwidth]{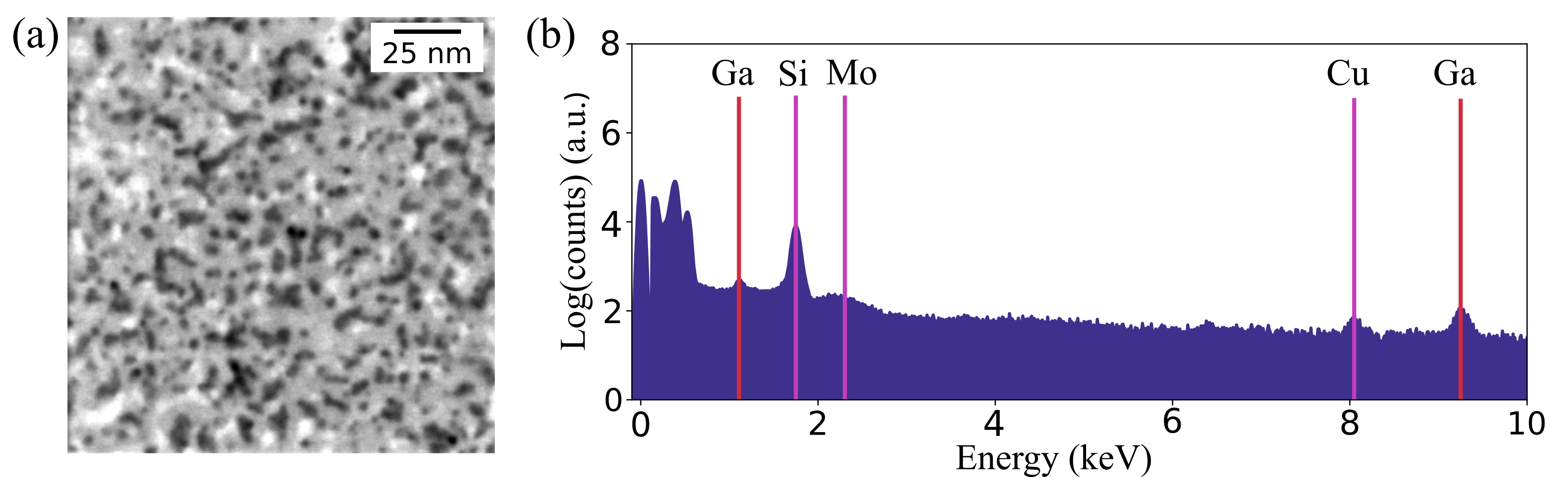}
  \caption{(a) HAADF-STEM image of multilayer hBN sample exposed to neutrals from the electrostatically blanked Ga LMIS for 20 hours. Dark contrast corresponds to pore-like structures. (b) Corresponding STEM-EDS signal integrated over this region with counts plotted on a log scale. Additional labeling of low-energy X-ray peaks in Supplementary Fig.~S4.}
  \label{fig:EDS}
\end{figure*}

From surveys of the pore structures over larger fields of view, we estimate the areal spread of neutral Ga clusters on the sample to be at least \qty{2e4}{\micro m^2}. Given a pore density of \qty{890}{pores/\micro m^2} for an exposure time of 20 minutes (from Fig.~\ref{fig:low mag}(a)), we estimate a neutral cluster flux of $\sim$\qty{1.5e4}{clusters/s}. The emission current of the LMIS was set to \qty{2}{\micro A}, which corresponds to \qty{1.25e13}{ions/s}. Therefore, in comparison with the ion flux, the flux of neutral clusters is very low, in agreement with previous studies~\citep{Mair1981MassSources,Mair1979Gallium-field-ionAnodes}. 

With regard to the kinetic energy of the neutral clusters that formed the pores, this will depend on whether the clusters were emitted as neutrals from the source, or were emitted charged and then subsequently neutralized. Post-emission neutralization could have occurred before, during or after transit through the extraction and accelerator fields. Given the size and depth of the impact sites (analyzed further below), cluster energies in the keV range seem most likely~\citep{Aoki2010,Anders2005}, pointing to clusters that were emitted charged and experienced the extraction and acceleration fields before neutralization. With an acceleration potential $V_{acc}$ of \qty{30}{kV} and assuming a charge state of $q=1e$, this would result in a maximum kinetic energy of $qV_{acc}=\qty{30}{keV}$. A doubly charged cluster ($q=2e$) would be accelerated to \qty{60}{keV}, etc. The final landing energy on the target will depend on energy transfer during neutralization and any other collisions in the column.

To compare neutral cluster emission from different Ga sources, a newly installed LMIS was tested on the same FIB microscope under the same conditions. TEM images of exposed samples again reveal \qty{1}{}--\qty{3}{nm} pore-like features (see Supplementary Information Fig.~S3). 
For the second LMIS, the density of these features was ~600 times lower, indicating a reduced neutral Ga cluster flux. It is likely that different sources will generate different amounts of Ga clusters depending on factors such as the amount of liquid Ga coating the tip and the dimensions of the Taylor cone.

\subsection{Detecting implanted Ga in the hBN sample}

Our findings suggest that neutral Ga clusters are responsible for the observed pore-like features in the exposed samples. It follows that in these samples one could also expect a degree of Ga implantation. This is now investigated using elemental mapping by STEM-EDS.  

Figure~\ref{fig:EDS}(a) shows a high angle annular dark field (HAADF) STEM image and \ref{fig:EDS}(b) the corresponding STEM-EDS signal integrated over the full field of view for a multilayer hBN sample (approximately \qty{15}{nm} thick) that had been exposed to the ion-beam-blanked Ga LMIS for 20 hours. The K$_\alpha$ and L$_\alpha$ X-ray peaks for Ga (at \qty{9.25}{keV} and \qty{1.10}{keV}, respectively) are labeled. Other elements presenting in the spectra include Si, Mo, and Cu, which can be attributed to background signal due to electron scattering from the substrate, sample holder, and microscope column components. Further discussion and labeling of the low-energy EDS peaks (where those for B and N are found) is given in the Supplementary Information (Fig.~S4(a)).

The integrated EDS signal clearly shows distinct Ga peaks, although quantification analysis reveals a Ga concentration of just \qty{0.02}{\text{at.\%}}. As a result, the computed elemental map for Ga (shown in Supplementary Fig.~S4(b)) is very noisy, since the signal per pixel is much lower and comparable to background noise. However, if the Ga clusters predominantly formed through-pores (as we later show to be the case for a \qty{7}{nm} thick hBN sample), then it would indeed be expected that most of the Ga passes through with minimal implantation. 

\subsection{Investigating nanopore size distributions}

\begin{figure*}[]
\centering
\includegraphics[width = 0.95\textwidth]{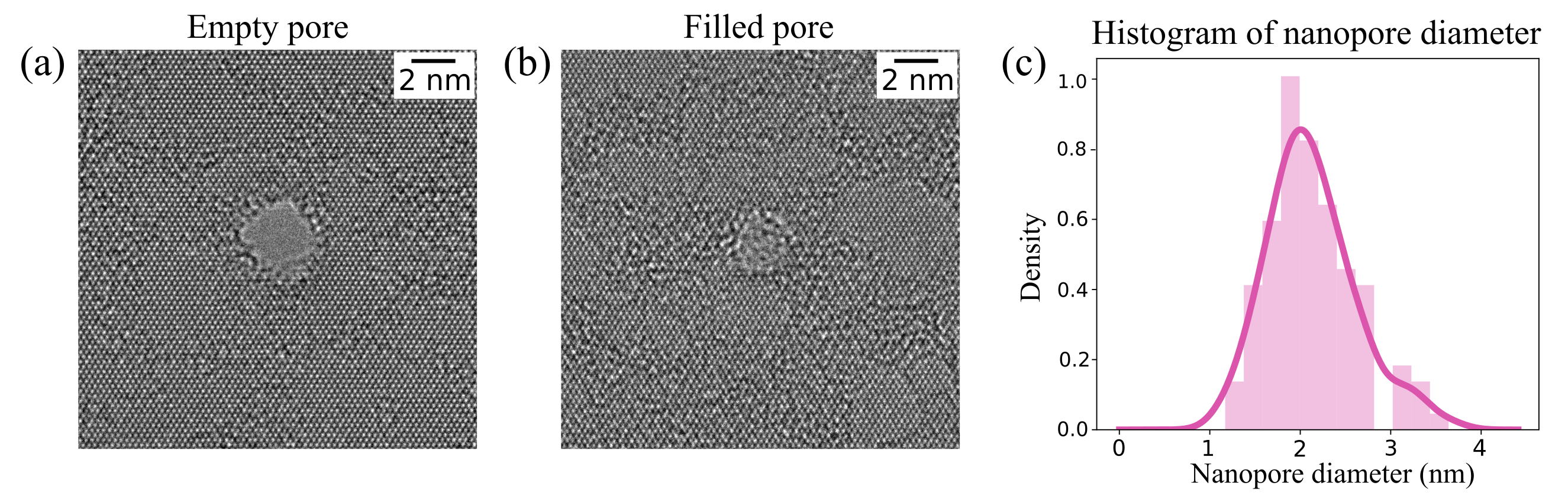}
  \caption{(a) HR-TEM image of a single nanopore formed in the multilayer hBN sample showing amorphous contamination around the perimeter. (b) HR-TEM image of another nanopore in the same sample that appears to be filled with amorphous material. (c) Histogram of nanopore diameters calculated by measuring 106 nanopores. The curve shows a kernel density estimate overlay.}
  \label{fig:Advanced characterize}
\end{figure*}

HR-TEM images of individual nanopore structures that formed in the hBN sample are shown in Figs.~\ref{fig:Advanced characterize}(a) and (b). Both images show patches of grainy texture due to an amorphous contamination layer on the hBN, which is known to be prevalent on 2D material samples~\citep{Schweizer2020MechanicalMicroscopy,Dyck2024YourClean}. In fact, hydrocarbon contamination is widely present on samples in general, but since the layer can be relatively thin, it does not necessarily show strong contrast, especially on thicker samples. In Fig.~\ref{fig:Advanced characterize}(a), the nanopore appears to have amorphous material around its perimeter, but the pore itself is open. In comparison, the nanopore in Fig.~\ref{fig:Advanced characterize}(b) is obscured by amorphous material. 

Fig.~\ref{fig:Advanced characterize}(c) shows a histogram of nanopore diameters obtained from measurements of 106 pores in the 20 minute exposed hBN sample. The overlay shows the corresponding kernel density estimate plot. The calculated mean nanopore diameter for this dataset was \qty{2.13}{nm} with a standard deviation of \qty{0.48}{nm}. The size distribution appears to be asymmetric, possibly indicating a multi-modal distribution with a second minor peak at \qty{\sim3.3}{nm}. As mentioned previously, two Ga LMISs were tested. For the second source (measuring 75 nanopores in a separate 20 minute exposed sample), a mean nanopore diameter of \qty{2.05}{nm} $\pm$ \qty{0.45}{nm} was calculated. The corresponding histogram is shown in Supplementary Fig.~S3(c).
For both sources tested, the nanopore size distributions are closely matched.

The sizes of the nanopores will depend on the size and kinetic energy of the Ga neutral clusters that formed them. 
Assuming minimal damage in the lateral direction, a \qty{2}{nm} nanopore could be formed by a Ga cluster of around the same diameter. However, given the concentrated energy deposition, there may in fact be significant peripheral structural damage meaning that a much smaller cluster could create a nanopore of this size \citep{Aoki2010}. Previous work investigating charged clusters from Ga LMIS by mass spectrometry observed a range of cluster sizes from 2--100 atoms, finding preferences for certain sizes~\citep{Barr1987GalliumSource,Saito1989AtomsSource,Sakaguchi1991AluminumSource,Bhaskar1987EvidenceSource}. 
This may explain the possibly bimodal (or higher order) distribution in nanopore sizes seen in Fig.~\ref{fig:Advanced characterize}(c).

\subsection{Depth sectioning of nanopores and spectroscopic analysis}

\begin{figure*}[]
\centering
\includegraphics[width = 0.7\textwidth]{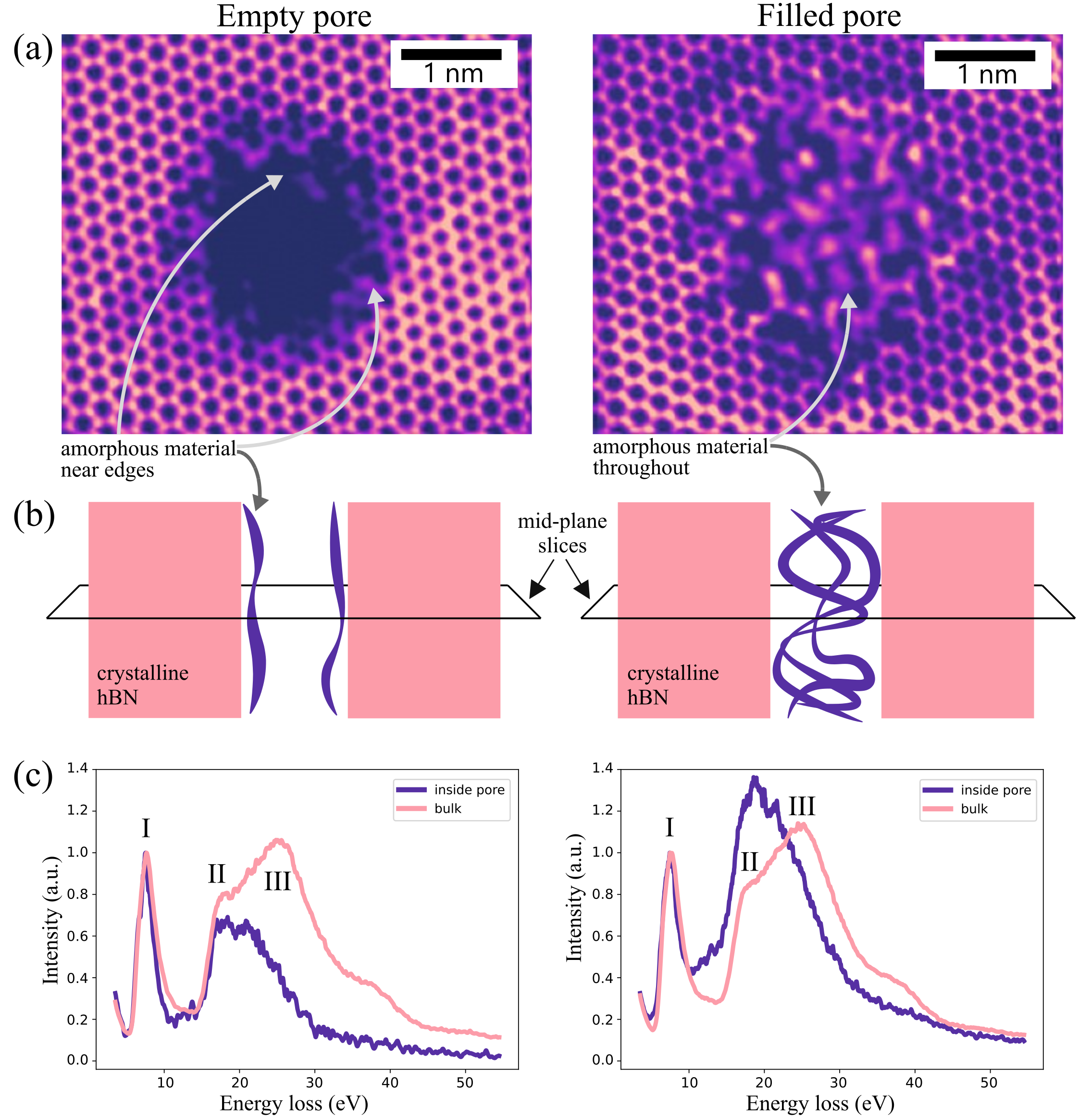}
  \caption{(a) Middle slices from two ptychographic depth-sectioning reconstructions for an empty pore (left) and a filled pore (right). (b) Corresponding vertical structure schematics with the intersecting mid-plane slices outlined in black. (c) Correlative STEM-EELS low-loss signals extracted from the regions inside the respective empty and filled pores, compared to the signals extracted from the bulk hBN around each pore in each case. Labels I, II, and III indicate plasmons and interband transitions from bulk hBN, discussed further in the main text.}
  \label{fig:Ptycho}
\end{figure*}

Figure~\ref{fig:Ptycho}(a) shows multislice electron ptychography results for two nanopores in \qty{7}{nm} thick hBN, in each case showing a slice corresponding to the depth-center of the membrane.
HAADF-STEM images and further slices from the reconstruction are shown in the Supplementary Figs.~S5--S8.
From the ptychography, we find that the nanopore on the left is open throughout the depth of the membrane, whereas the one on the right is filled with amorphous material throughout its depth. This is shown schematically in Fig.~\ref{fig:Ptycho}(b).

The depth sectioning analysis also reveals that the top and bottom openings of the nanopores are approximately the same size, with near-parallel sidewalls. There is also no discernible redeposited material around the top perimeter of the pores. (See Supplementary Figs.~S6--S8 and Supplementary Videos 1 and 2.) 
These results indicate that the Ga clusters had sufficient kinetic energy to pass through the membrane entirely, rather than solely depositing energy at the surface, which would have created wider openings on the top side with crater-like accumulation around the edge~\citep{Aoki2010}.  

The amorphous material filling the nanopore on the right of Fig.~\ref{fig:Ptycho}(a) could be the same amorphous (hydrocarbon) contamination observed on the surface of these samples by HR-TEM, which has migrated into the pore. For the empty pore on the left, we observe a thin amorphous layer (\qty{\sim0.4}{nm} thick) along the sidewalls. This could also be migrated hydrocarbon contamination. A degree of amorphization of the hBN along the edges of the pore due to energy deposition from the transiting Ga cluster can also be expected~\citep{Anders2005}. 

In order to analyze the amorphous material in more detail, we performed correlative low-loss STEM-EELS for the same two nanopores investigated above by ptychography, as shown in Fig.~\ref{fig:Ptycho}(c). Spectra have been extracted for the interior (`inside pore') and exterior (`bulk') regions, normalized to the first peak. The main peaks in the hBN spectra are labeled. Peaks I and III correspond to the $\pi$ and $\pi + \sigma$ plasmons of bulk hBN (at \qty{\sim7.5}{eV} and \qty{25}{eV}, respectively)~\citep{Fossard2017}. Peak II (at \qty{\sim17.5}{eV}) manifests as a shoulder to the $\pi + \sigma$ plasmon and has previously been attributed to interband transitions~\citep{Arnaud2006}. 

In the case of the spectrum corresponding to the inside of the empty pore (on the left), peaks I and II are preserved, but peak III has significantly reduced in intensity. While the tails of the beam will clip the hBN when close to the edge, the spectral features observed here can largely be attributed to long-range probing of delocalized charge oscillations by the tightly focused electron beam as it scans over the vacuum space inside the pore. This is also known as aloof mode EELS~\citep{Krivanek2014}. In aloof mode, the intensity of the energy-loss peaks decreases with energy-loss value~\citep{Rez2016,Crozier2017}, supporting the progressive decrease in peak intensities observed in our vacuum measurement.

For the spectrum corresponding to the inside of the filled pore (on the right), we see that the hBN peak I is again preserved, presumably also due to delocalization effects. However, the main difference is that the angular shape formed by hBN peaks II and III has disappeared, with a broad peak centered at \qty{\sim20}{eV} rising in its place. Elsewhere a reminiscent transitioning behavior in the low-loss spectrum has been observed for a thin hBN sample that was milled with the STEM electron beam in situ~\citep{Clark2019}. As the hBN was gradually removed, the hBN peaks we label as II and III transitioned into a very similar broader peak. We suggest that this peak could be the $\pi + \sigma$ plasmon from amorphous hydrocarbon contamination~\citep{Lifshitz2003,Sola2009}, which in the above-cited work could arise due to beam-induced deposition of mobile hydrocarbons~\citep{Dyck2024YourClean}---and in our study, can be due to the migration of surface contamination into the pore. 
Amorphization of the hBN itself can also be expected, both due to the electron beam milling described, and as a result of the gallium cluster bombardment of our work. Such amorphization would be expected to manifest as a general broadening and energy shift of the hBN plasmon peaks. However, peak I for the spectrum inside the pore appears to show neither broadening nor shift, suggesting that any amorphous hBN inside the pore that may contribute to this peak is negligible compared to the delocalized signal emanating from the crystalline hBN outside. 

In summary, the low-loss EELS mapping corroborates the ptychography result that the filled pores contain amorphous material. The EELS data suggest that this amorphous material is hydrocarbon contamination, although the possibility of amorphous hBN cannot be discounted. 

\subsection{Expansion and cleaning of nanopores}

\begin{figure}[]
\centering
\includegraphics[height = 8 cm]{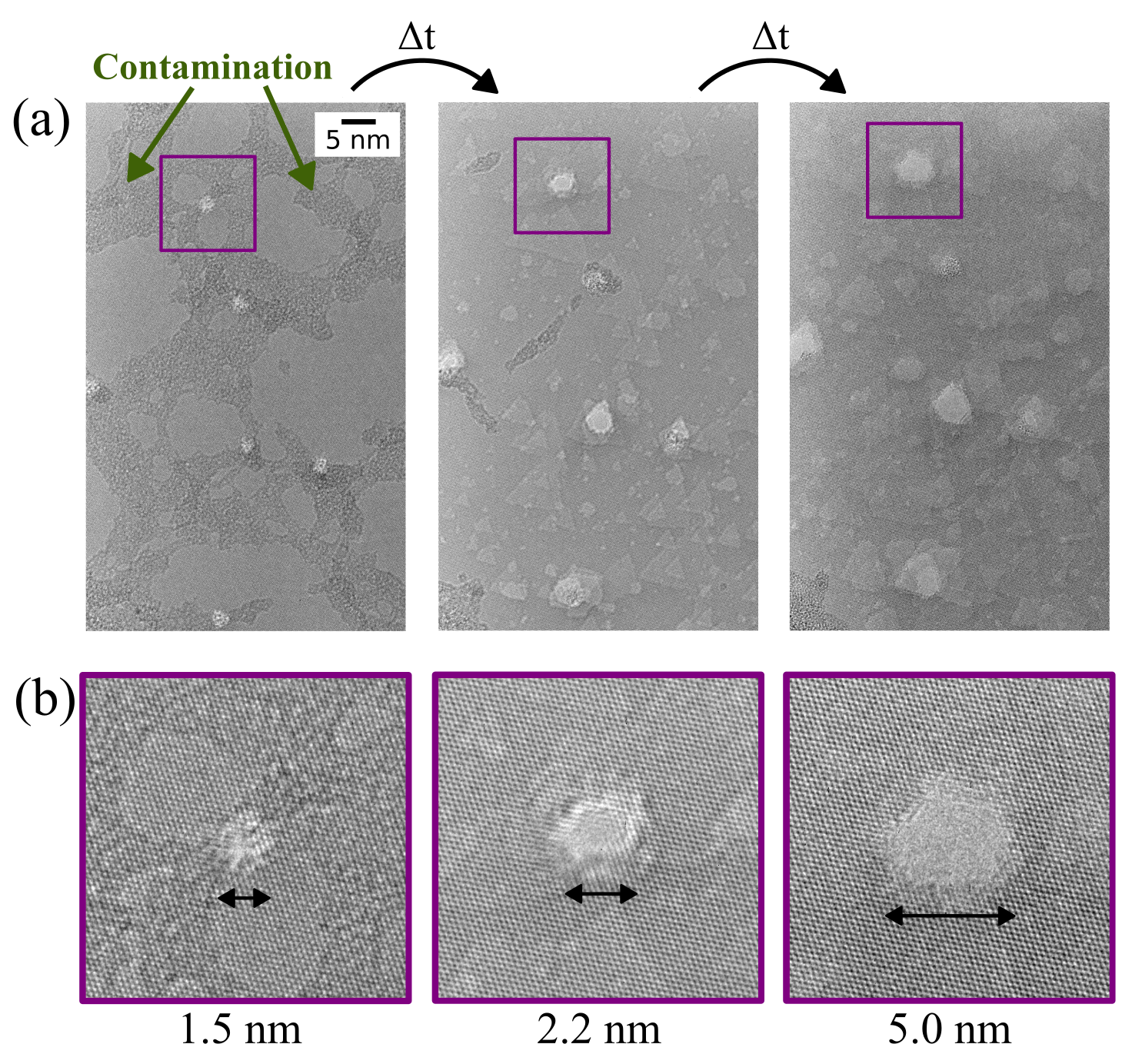}
  \caption{(a) HR-TEM inspection of nanopores in hBN that had been subjected to electron irradiation at a dose rate of \SI{\sim4e6}{e^-/nm^2/s} to remove hydrocarbon contamination and expand pores to larger sizes. ($\Delta$t = \qty{150}{s}). (b) Zoomed-in images of the highlighted nanopore in (a) showing its expansion over time.}
  \label{fig:nanopore growth}
\end{figure}

Although Ga cluster irradiation from the LMIS can damage delicate samples in an unintentional manner, the small size and narrow size distribution of the nanopores formed in the thin freestanding membranes investigated here make them interesting from an application standpoint. For example, we propose that Ga clusters emitted from a LMIS can be used to fabricate nanopores for nanofluidics applications such as molecular or nanoparticle separation, which requires a large number of consistently sized pores for high efficiency and selectivity. By exposing samples to the electrostatically blanked Ga LMIS for varying amounts of time, the nanopore density can be controlled. 
However, in order to be useful for transport applications, pores that are filled with amorphous material need to be unclogged. Furthermore, tuning of the pore size to suit the particular separation experiment to be performed would be highly beneficial. 

Here we show that both pore unclogging and expansion can be achieved by electron irradiation in a TEM. The results for Ga-cluster-induced nanopores in an hBN sample are shown in Fig.~\ref{fig:nanopore growth}, for which \qty{80}{keV} electron irradiation in parallel beam mode was employed. In Fig.~\ref{fig:nanopore growth}(a) we see TEM images acquired at the beginning of the process and at two successive time points of \qty{150}{s} and \qty{300}{s} (corresponding to total accumulated doses of \SI{\sim6e8}{e^-/nm^2} and \SI{\sim1.2e9}{e^-/nm^2}, respectively). Zoomed-in views tracking an individual nanopore from each time point are shown in Fig.~\ref{fig:nanopore growth}(b). The corresponding full field-of-view HR-TEM images are shown in Supplementary Fig.~S9.

As the exposure time increases, the amorphous contamination coverage of the sample is reduced. At the same time, the nanopores are enlarged. This is clearly seen in the zoomed-in views, where the tracked nanopore is unclogged and increases in size from \SIrange{1.5}{5}{nm}.
All nanopores in the field of view appear to grow at a similar rate. 
The mechanism responsible for material removal under the electron beam involves a combination of elastic interactions (knock-on damage) and inelastic effects (radiolysis, etc.)~\citep{Egerton2004}. After the first few seconds of electron irradiation, the emergence of faint triangular-shaped features is also observed. These correspond to gradual exfoliation of the bulk hBN membrane under the electron beam, with selective sputtering of one atomic species over the other driving the triangular shape~\citep{Meyer2009}. 

\section{Conclusions}

We have investigated the effect of exposing freestanding membrane samples to a beam-blanked Ga LMIS. We find that nanopores of diameter \qty{1}{}--{3}{nm} are formed, which we attribute to individual impacts from neutral Ga clusters that evade the electrostatic (i.e.~charge-based) beam blanking system. Given the areal coverage of the impact sites, the flux of neutral Ga clusters is estimated to be \SI{\sim1.5E4}{clusters/s}. After $\sim$20 minutes of exposure, nanopores that are on average spaced tens of nanometers apart were found. While sparse, this effect could still be detrimental for samples that must remain pristine except in regions that are to be intentionally processed by the FIB. 

HR-STEM, STEM-EDS, multislice electron ptychography and STEM-EELS enabled detailed characterization of the nanopore structure and composition. For membrane samples of thickness \qty{7}{nm}, we find that the pores constitute through-holes with near-parallel sidewalls. In future work, it will be interesting to work with thicker films to determine the maximum channel depth and hence estimate the landing energy of the clusters. 

Most nanopores are clogged with amorphous material, while a few are open. Nevertheless, the narrow size distribution of the nanopores makes them promising candidates for nanofluidics applications. We show that high-dose electron irradiation in the TEM can be used to controllably remove the amorphous clogging material and grow the pores, providing a unique way to tune permeability for size-selective membrane applications. 

\section{Competing interests}
No competing interest is declared.

\section{Author contributions statement}

D.O.B. and F.I.A. conceived the experiments,  D.O.B. and S.M.R. conducted the experiments, all authors analyzed the results,  D.O.B. and F.I.A. wrote the manuscript draft, all authors reviewed the manuscript.

\section{Acknowledgments}
This work was funded in part by NSF Award No.\ 2110924. D.O.B. also acknowledges funding from the Department of Defense through the National Defense Science \& Engineering Graduate (NDSEG) Fellowship Program. 
S.M.R. and C.O. acknowledge support from the U.S. Department of Energy Early Career Research Award program.
The irradiation experiments were performed at the Biomolecular Nanotechnology Center, a core facility of the California Institute for Quantitative Biosciences.
Work at the Molecular Foundry was supported by the Office of Science, Office of Basic Energy Sciences, of the U.S. Department of Energy under Contract No.\ DE-AC02-05CH11231.
This research also used resources of the National Energy Research Scientific Computing Center (NERSC), a Department of Energy Office of Science User Facility using NERSC award BES-ERCAP0028035.

\bibliographystyle{unsrtnat}
\bibliography{bib}

\begin{thebibliography}{39}
\providecommand{\natexlab}[1]{#1}
\providecommand{\url}[1]{\texttt{#1}}
\expandafter\ifx\csname urlstyle\endcsname\relax
  \providecommand{\doi}[1]{doi: #1}\else
  \providecommand{\doi}{doi: \begingroup \urlstyle{rm}\Url}\fi

\bibitem[Höflich et~al.(2023)Höflich, Hobler, Allen, Wirtz, Rius,
  McElwee-White, Krasheninnikov, Schmidt, Utke, Klingner, Osenberg, Córdoba,
  Djurabekova, Manke, Moll, Manoccio, De~Teresa, Bischoff, Michler, De~Castro,
  Delobbe, Dunne, Dobrovolskiy, Frese, Gölzhäuser, Mazarov, Koelle, Möller,
  Pérez-Murano, Philipp, Vollnhals, and Hlawacek]{Hoflich2023}
Katja Höflich, Gerhard Hobler, Frances~I Allen, Tom Wirtz, Gemma Rius, Lisa
  McElwee-White, Arkady~V Krasheninnikov, Matthias Schmidt, Ivo Utke, Nico
  Klingner, Markus Osenberg, Rosa Córdoba, Flyura Djurabekova, Ingo Manke,
  Philip Moll, Mariachiara Manoccio, José~María De~Teresa, Lothar Bischoff,
  Johann Michler, Olivier De~Castro, Anne Delobbe, Peter Dunne, Oleksandr~V
  Dobrovolskiy, Natalie Frese, Armin Gölzhäuser, Paul Mazarov, Dieter Koelle,
  Wolfhard Möller, Francesc Pérez-Murano, Patrick Philipp, Florian Vollnhals,
  and Gregor Hlawacek.
\newblock Roadmap for focused ion beam technologies.
\newblock \emph{Appl. Phys. Rev.}, 10\penalty0 (4):\penalty0 041311, 2023.

\bibitem[Mayer et~al.(2007)Mayer, Giannuzzi, Kamino, and Michael]{Mayer2007}
Joachim Mayer, Lucille~A Giannuzzi, Takeo Kamino, and Joseph Michael.
\newblock {TEM} sample preparation and {FIB}-induced damage.
\newblock \emph{MRS Bulletin}, 32\penalty0 (5):\penalty0 400--407, 2007.

\bibitem[Berger et~al.(2023)Berger, Premaraj, Ravelli, Knoops, López-Iglesias,
  and Peters]{Berger2023}
Casper Berger, Navya Premaraj, Raimond B~G Ravelli, Kèvin Knoops, Carmen
  López-Iglesias, and Peter~J Peters.
\newblock Cryo-electron tomography on focused ion beam lamellae transforms
  structural cell biology.
\newblock \emph{Nat. Methods}, 20\penalty0 (4):\penalty0 499--511, 2023.

\bibitem[Scheffer et~al.(2020)Scheffer, Xu, Januszewski, Lu, Takemura,
  Hayworth, Huang, Shinomiya, Maitlin-Shepard, Berg, Clements, Hubbard, Katz,
  Umayam, Zhao, Ackerman, Blakely, Bogovic, Dolafi, Kainmueller, Kawase,
  Khairy, Leavitt, Li, Lindsey, Neubarth, Olbris, Otsuna, Trautman, Ito, Bates,
  Goldammer, Wolff, Svirskas, Schlegel, Neace, Knecht, Alvarado, Bailey,
  Ballinger, Borycz, Canino, Cheatham, Cook, Dreher, Duclos, Eubanks,
  Fairbanks, Finley, Forknall, Francis, Hopkins, Joyce, Kim, Kirk, Kovalyak,
  Lauchie, Lohff, Maldonado, Manley, McLin, Mooney, Ndama, Ogundeyi, Okeoma,
  Ordish, Padilla, Patrick, Paterson, Phillips, Phillips, Rampally, Ribeiro,
  Robertson, Rymer, Ryan, Sammons, Scott, Scott, Shinomiya, Smith, Smith,
  Smith, Sobeski, Suleiman, Swift, Takemura, Talebi, Tarnogorska, Tenshaw,
  Tokhi, Walsh, Yang, Horne, Li, Parekh, Rivlin, Jayaraman, Costa, Jefferis,
  Ito, Saalfeld, George, Meinertzhagen, Rubin, Hess, Jain, and
  Plaza]{Scheffer2020}
Louis~K Scheffer, C~Shan Xu, Michal Januszewski, Zhiyuan Lu, Shin-Ya Takemura,
  Kenneth~J Hayworth, Gary~B Huang, Kazunori Shinomiya, Jeremy Maitlin-Shepard,
  Stuart Berg, Jody Clements, Philip~M Hubbard, William~T Katz, Lowell Umayam,
  Ting Zhao, David Ackerman, Tim Blakely, John Bogovic, Tom Dolafi, Dagmar
  Kainmueller, Takashi Kawase, Khaled~A Khairy, Laramie Leavitt, Peter~H Li,
  Larry Lindsey, Nicole Neubarth, Donald~J Olbris, Hideo Otsuna, Eric~T
  Trautman, Masayoshi Ito, Alexander~S Bates, Jens Goldammer, Tanya Wolff,
  Robert Svirskas, Philipp Schlegel, Erika Neace, Christopher~J Knecht,
  Chelsea~X Alvarado, Dennis~A Bailey, Samantha Ballinger, Jolanta~A Borycz,
  Brandon~S Canino, Natasha Cheatham, Michael Cook, Marisa Dreher, Octave
  Duclos, Bryon Eubanks, Kelli Fairbanks, Samantha Finley, Nora Forknall,
  Audrey Francis, Gary~Patrick Hopkins, Emily~M Joyce, Sungjin Kim, Nicole~A
  Kirk, Julie Kovalyak, Shirley~A Lauchie, Alanna Lohff, Charli Maldonado,
  Emily~A Manley, Sari McLin, Caroline Mooney, Miatta Ndama, Omotara Ogundeyi,
  Nneoma Okeoma, Christopher Ordish, Nicholas Padilla, Christopher~M Patrick,
  Tyler Paterson, Elliott~E Phillips, Emily~M Phillips, Neha Rampally, Caitlin
  Ribeiro, Madelaine~K Robertson, Jon~Thomson Rymer, Sean~M Ryan, Megan
  Sammons, Anne~K Scott, Ashley~L Scott, Aya Shinomiya, Claire Smith, Kelsey
  Smith, Natalie~L Smith, Margaret~A Sobeski, Alia Suleiman, Jackie Swift,
  Satoko Takemura, Iris Talebi, Dorota Tarnogorska, Emily Tenshaw, Temour
  Tokhi, John~J Walsh, Tansy Yang, Jane~Anne Horne, Feng Li, Ruchi Parekh,
  Patricia~K Rivlin, Vivek Jayaraman, Marta Costa, Gregory~Sxe Jefferis, Kei
  Ito, Stephan Saalfeld, Reed George, Ian~A Meinertzhagen, Gerald~M Rubin,
  Harald~F Hess, Viren Jain, and Stephen~M Plaza.
\newblock A connectome and analysis of the adult drosophila central brain.
\newblock \emph{Elife}, 9:\penalty0 e57443, 2020.

\bibitem[Hollenbach et~al.(2022)Hollenbach, Klingner, Jagtap, Bischoff, Fowley,
  Kentsch, Hlawacek, Erbe, Abrosimov, Helm, Berencén, and
  Astakhov]{Hollenbach2022}
Michael Hollenbach, Nico Klingner, Nagesh~S Jagtap, Lothar Bischoff, Ciarán
  Fowley, Ulrich Kentsch, Gregor Hlawacek, Artur Erbe, Nikolay~V Abrosimov,
  Manfred Helm, Yonder Berencén, and Georgy~V Astakhov.
\newblock Wafer-scale nanofabrication of telecom single-photon emitters in
  silicon.
\newblock \emph{Nat. Commun.}, 13\penalty0 (1):\penalty0 7683, 2022.

\bibitem[Audinot et~al.(2021)Audinot, Philipp, De~Castro, Biesemeier, Hoang,
  and Wirtz]{Audinot2021}
Jean-Nicolas Audinot, Patrick Philipp, Olivier De~Castro, Antje Biesemeier,
  Quang~Hung Hoang, and Tom Wirtz.
\newblock Highest resolution chemical imaging based on secondary ion mass
  spectrometry performed on the helium ion microscope.
\newblock \emph{Rep. Prog. Phys.}, 84\penalty0 (10):\penalty0 105901, 2021.

\bibitem[Taylor(1964)]{Taylor1964DisintegrationField}
Geoffrey Taylor.
\newblock {Disintegration of water drops in an electric field}.
\newblock \emph{Proc. R. Soc. Lond. A}, 280\penalty0 (1382):\penalty0 383--397,
  1964.

\bibitem[Gomer(1979)]{Gomer1979AppliedSources}
R~Gomer.
\newblock {Applied Physics On the Mechanism of Liquid Metal Electron and Ion
  Sources}.
\newblock \emph{Appl. Phys.}, 19\penalty0 (4):\penalty0 365--375, 1979.

\bibitem[Mair and Von~Engel(1981)]{Mair1981MassSources}
G~L~R Mair and A~Von~Engel.
\newblock {Mass transport in liquid gallium ion beam sources}.
\newblock \emph{J. Phys. D: Appl. Phys}, 14\penalty0 (9):\penalty0 1721--1729,
  1981.

\bibitem[Mair and Von~Engel(1979)]{Mair1979Gallium-field-ionAnodes}
G~L~R Mair and A~Von~Engel.
\newblock {Gallium-field-ion emission from liquid point anodes}.
\newblock \emph{Journal of Applied Physics}, 50\penalty0 (9):\penalty0
  5592--5595, 1979.

\bibitem[Kagarice et~al.(2017)Kagarice, Otis, and William~Parker]{Kagarice2017}
Kevin Kagarice, Charles Otis, and N~William~Parker.
\newblock Chicane blanker assemblies for charged particle beam systems and
  methods of using the same, 2017.
\newblock U.S. Patent 9767984.

\bibitem[D'Cruz et~al.(1985)D'Cruz, Pourrezaei, and
  Wagner]{DCruz1985IonSources}
C~D'Cruz, K~Pourrezaei, and A~Wagner.
\newblock {Ion cluster emission and deposition from liquid gold ion sources}.
\newblock \emph{Journal of Applied Physics}, 58\penalty0 (7):\penalty0
  2724--2730, 1985.

\bibitem[Wagner et~al.(1981)Wagner, Venkatesan, Petroff, and
  Barr]{Wagner1981DROPLETSOURCES.}
A~Wagner, T~Venkatesan, P~M Petroff, and D~Barr.
\newblock Droplet emission in liquid metal ion sources.
\newblock \emph{Journal of Vacuum Science {\&} Technology}, 19\penalty0
  (4):\penalty0 1186--1189, 1981.

\bibitem[Culbertson et~al.(1979)Culbertson, Robertson, Kuk, and
  Sakurai]{Culbertson1979ATOM-PROBESOURCE.}
Robert~J Culbertson, G~H Robertson, Y~Kuk, and T~Sakurai.
\newblock Atom-probe field-ion microscopy of a high intensity gallium ion
  source.
\newblock \emph{Journal of Vacuum Science {\&} Technology}, 17\penalty0
  (1):\penalty0 203--206, 1979.

\bibitem[Barr(1987)]{Barr1987GalliumSource}
D~L Barr.
\newblock {Gallium clusters from a liquid metal ion source}.
\newblock \emph{Journal of Vacuum Science {\&} Technology B: Microelectronics
  and Nanometer Structures}, 5\penalty0 (1):\penalty0 184--189, 1987.

\bibitem[Saito et~al.(1989)Saito, Minami, Ishida, and
  Noda]{Saito1989AtomsSource}
Y~Saito, K~Minami, T~Ishida, and T~Noda.
\newblock {Abundance of Na cluster ions ejected from a liquid metal ion
  source}.
\newblock \emph{Z. Phys. D Atoms, Molecules and Clusters}, 11\penalty0
  (1):\penalty0 87--91, 1989.

\bibitem[Sakaguchi et~al.(1991)Sakaguchi, Mihama, and
  Saito]{Sakaguchi1991AluminumSource}
Kenji Sakaguchi, Kazuhiro Mihama, and Yahachi Saito.
\newblock {Aluminum cluster ions generated by a liquid metal ion source}.
\newblock \emph{Journal of Applied Physics}, 70\penalty0 (9):\penalty0
  5049--5053, 1991.

\bibitem[Bhaskar et~al.(1987)Bhaskar, Frueholz, Klimcak, and
  Cook]{Bhaskar1987EvidenceSource}
N~D Bhaskar, R~P Frueholz, C~M Klimcak, and R~A Cook.
\newblock {Evidence of electronic shell structure in Rb$_N$$^+$ (N = 1-100)
  produced in a liquid-metal ion source}.
\newblock \emph{Phys. Rev. B}, 36\penalty0 (8):\penalty0 4418--4421, 1987.

\bibitem[Castellanos-Gomez et~al.(2014)Castellanos-Gomez, Buscema, Molenaar,
  Singh, Janssen, van~der Zant, and Steele]{Castellanos-Gomez2014}
Andres Castellanos-Gomez, Michele Buscema, Rianda Molenaar, Vibhor Singh,
  Laurens Janssen, Herre S~J van~der Zant, and Gary~A Steele.
\newblock Deterministic transfer of two-dimensional materials by all-dry
  viscoelastic stamping.
\newblock \emph{2D Mater.}, 1\penalty0 (1):\penalty0 011002, 2014.

\bibitem[Ophus(2019)]{Ophus2019}
Colin Ophus.
\newblock Four-dimensional scanning transmission electron microscopy
  ({4D}-{STEM}): From scanning nanodiffraction to ptychography and beyond.
\newblock \emph{Microsc. Microanal.}, 25\penalty0 (3):\penalty0 563--582, 2019.

\bibitem[Chen et~al.(2021)Chen, Jiang, Shao, Holtz, Odstr{\v{c}}il,
  Guizar-Sicairos, Hanke, Ganschow, Schlom, and Muller]{chen2021electron}
Zhen Chen, Yi~Jiang, Yu-Tsun Shao, Megan~E Holtz, Michal Odstr{\v{c}}il, Manuel
  Guizar-Sicairos, Isabelle Hanke, Steffen Ganschow, Darrell~G Schlom, and
  David~A Muller.
\newblock Electron ptychography achieves atomic-resolution limits set by
  lattice vibrations.
\newblock \emph{Science}, 372\penalty0 (6544):\penalty0 826--831, 2021.

\bibitem[Ribet et~al.(2024)Ribet, Varnavides, Pedroso, Cohen, Ercius, Scott,
  and Ophus]{ribet2024uncovering}
Stephanie~M Ribet, Georgios Varnavides, Cassio Pedroso, Bruce~E Cohen, Peter
  Ercius, Mary~C Scott, and Colin Ophus.
\newblock Uncovering the three-dimensional structure of upconverting
  core--shell nanoparticles with multislice electron ptychography.
\newblock \emph{Applied Physics Letters}, 124\penalty0 (24):\penalty0 240601,
  2024.

\bibitem[Varnavides et~al.(2023)Varnavides, Ribet, Zeltmann, Yu, Savitzky,
  Byrne, Allen, Dravid, Scott, and Ophus]{varnavides2023iterative}
Georgios Varnavides, Stephanie~M Ribet, Steven~E Zeltmann, Yue Yu, Benjamin~H
  Savitzky, Dana~O Byrne, Frances~I Allen, Vinayak~P Dravid, Mary~C Scott, and
  Colin Ophus.
\newblock Iterative phase retrieval algorithms for scanning transmission
  electron microscopy.
\newblock \emph{arXiv preprint arXiv:2309.05250}, 2023.

\bibitem[de~la Peña et~al.(2023)]{HyperSpy}
Francisco de~la Peña et~al.
\newblock hyperspy/hyperspy: v2.0rc0, November 2023.

\bibitem[Aoki et~al.(2010)Aoki, Seki, and Matsuo]{Aoki2010}
Takaaki Aoki, Toshio Seki, and Jiro Matsuo.
\newblock Molecular dynamics simulations for gas cluster ion beam processes.
\newblock \emph{Vacuum}, 84\penalty0 (8):\penalty0 994--998, 2010.

\bibitem[Anders and Urbassek(2005)]{Anders2005}
Christian Anders and Herbert~M. Urbassek.
\newblock Cluster-size dependence of ranges of 100e{V}/atom {A}u$_n$ clusters.
\newblock \emph{Nuclear Instruments and Methods in Physics Research Section B:
  Beam Interactions with Materials and Atoms}, 228\penalty0 (1):\penalty0
  57--63, 2005.

\bibitem[Thiruraman et~al.(2018)Thiruraman, Fujisawa, Danda, Das, Zhang,
  Bolotsky, Perea-L{\'o}pez, Nicola{\"\i}, Senet, Terrones, and
  Drndi{\'c}]{Thiruraman2018}
Jothi~Priyanka Thiruraman, Kazunori Fujisawa, Gopinath Danda, Paul~Masih Das,
  Tianyi Zhang, Adam Bolotsky, N{\'e}stor Perea-L{\'o}pez, Adrien Nicola{\"\i},
  Patrick Senet, Mauricio Terrones, and Marija Drndi{\'c}.
\newblock Angstrom-size defect creation and ionic transport through pores in
  single-layer {MoS$_2$}.
\newblock \emph{Nano Lett.}, 18\penalty0 (3):\penalty0 1651--1659, 2018.

\bibitem[Schweizer et~al.(2020)Schweizer, Dolle, Dasler, Abell{\'{a}}n, Hauke,
  Hirsch, and Spiecker]{Schweizer2020MechanicalMicroscopy}
Peter Schweizer, Christian Dolle, Daniela Dasler, Gonzalo Abell{\'{a}}n, Frank
  Hauke, Andreas Hirsch, and Erdmann Spiecker.
\newblock {Mechanical cleaning of graphene using in situ electron microscopy}.
\newblock \emph{Nature Communications}, 11\penalty0 (1):\penalty0 1743, 2020.

\bibitem[Dyck et~al.(2024)Dyck, Okmi, Xiao, Lei, Lupini, and
  Jesse]{Dyck2024YourClean}
Ondrej Dyck, Aisha Okmi, Kai Xiao, Sidong Lei, Andrew~R. Lupini, and Stephen
  Jesse.
\newblock {Your Clean Graphene is Still Not Clean}.
\newblock \emph{Advanced Materials Interfaces}, 12\penalty0 (1):\penalty0
  2400598, 2024.

\bibitem[Fossard et~al.(2017)Fossard, Sponza, Schué, Attaccalite, Ducastelle,
  Barjon, and Loiseau]{Fossard2017}
Frédéric Fossard, Lorenzo Sponza, Léonard Schué, Claudio Attaccalite,
  François Ducastelle, Julien Barjon, and Annick Loiseau.
\newblock Angle-resolved electron energy loss spectroscopy in hexagonal boron
  nitride.
\newblock \emph{Phys. Rev. B.}, 96\penalty0 (11):\penalty0 115304, 2017.

\bibitem[Arnaud et~al.(2006)Arnaud, Lebègue, Rabiller, and
  Alouani]{Arnaud2006}
B~Arnaud, S~Lebègue, P~Rabiller, and M~Alouani.
\newblock Huge excitonic effects in layered hexagonal boron nitride.
\newblock \emph{Phys. Rev. Lett.}, 96\penalty0 (2):\penalty0 026402, 2006.

\bibitem[Krivanek et~al.(2014)Krivanek, Lovejoy, Dellby, Aoki, Carpenter, Rez,
  Soignard, Zhu, Batson, Lagos, Egerton, and Crozier]{Krivanek2014}
Ondrej~L Krivanek, Tracy~C Lovejoy, Niklas Dellby, Toshihiro Aoki, R~W
  Carpenter, Peter Rez, Emmanuel Soignard, Jiangtao Zhu, Philip~E Batson,
  Maureen~J Lagos, Ray~F Egerton, and Peter~A Crozier.
\newblock Vibrational spectroscopy in the electron microscope.
\newblock \emph{Nature}, 514\penalty0 (7521):\penalty0 209--212, 2014.

\bibitem[Rez et~al.(2016)Rez, Aoki, March, Gur, Krivanek, Dellby, Lovejoy,
  Wolf, and Cohen]{Rez2016}
Peter Rez, Toshihiro Aoki, Katia March, Dvir Gur, Ondrej~L Krivanek, Niklas
  Dellby, Tracy~C Lovejoy, Sharon~G Wolf, and Hagai Cohen.
\newblock Damage-free vibrational spectroscopy of biological materials in the
  electron microscope.
\newblock \emph{Nat. Commun.}, 7\penalty0 (1):\penalty0 10945, 2016.

\bibitem[Crozier(2017)]{Crozier2017}
Peter~A Crozier.
\newblock Vibrational and valence aloof beam {EELS}: A potential tool for
  nondestructive characterization of nanoparticle surfaces.
\newblock \emph{Ultramicroscopy}, 180\penalty0 (6):\penalty0 104--114, 2017.

\bibitem[Clark et~al.(2019)Clark, Lewis, Haigh, and Vijayaraghavan]{Clark2019}
Nick Clark, Edward~A Lewis, Sarah~J Haigh, and Aravind Vijayaraghavan.
\newblock Nanometre electron beam sculpting of suspended graphene and hexagonal
  boron nitride heterostructures.
\newblock \emph{2D Mater.}, 6\penalty0 (2):\penalty0 025032, 2019.

\bibitem[Lifshitz(2003)]{Lifshitz2003}
Y~Lifshitz.
\newblock Pitfalls in amorphous carbon studies.
\newblock \emph{Diam. Relat. Mater.}, 12\penalty0 (2):\penalty0 130--140, 2003.

\bibitem[Solá et~al.(2009)Solá, Biaggi-Labiosa, Fonseca, Resto,
  Lebrón-Colón, and Meador]{Sola2009}
F~Solá, A~Biaggi-Labiosa, L~F Fonseca, O~Resto, M~Lebrón-Colón, and M~A
  Meador.
\newblock Field emission and radial distribution function studies of
  fractal-like amorphous carbon nanotips.
\newblock \emph{Nanoscale Res. Lett.}, 4\penalty0 (5):\penalty0 431--436, 2009.

\bibitem[Egerton et~al.(2004)Egerton, Li, and Malac]{Egerton2004}
R~F Egerton, P~Li, and M~Malac.
\newblock {Radiation damage in the TEM and SEM}.
\newblock \emph{Micron}, 35\penalty0 (6):\penalty0 399--409, 2004.

\bibitem[Meyer et~al.(2009)Meyer, Chuvilin, Algara-Siller, Biskupek, and
  Kaiser]{Meyer2009}
Jannik~C Meyer, Andrey Chuvilin, Gerardo Algara-Siller, Johannes Biskupek, and
  Ute Kaiser.
\newblock {Selective sputtering and atomic resolution imaging of atomically
  thin boron nitride membranes}.
\newblock \emph{Nano Letters}, 9\penalty0 (7):\penalty0 2683--2689, 2009.

\end{thebibliography}
\end{document}


\begin{frontmatter}

\title{\texorpdfstring{Supplementary Information:\\ \vspace{2mm} Neutral but Impactful: Gallium Cluster-Induced Nanopores from Beam-Blanked Gallium Ion Sources}}

\author[label1,label2,label3]{Dana O. Byrne}

\author[label3]{Stephanie M. Ribet}

\author[label3]{Karen C. Bustillo}

\author[label4]{Colin Ophus}

\author[label2,label3,label5]{\\Frances I. Allen}

\address[label1]{Department of Chemistry, UC Berkeley, Berkeley, CA 94720, USA}
\address[label2]{Department of Materials Science and Engineering, UC Berkeley, Berkeley, CA 94720, USA}
\address[label3]{National Center for Electron Microscopy, Molecular Foundry, LBNL, Berkeley, CA 94720, USA}
\address[label4]{Department of Materials Science and Engineering, Stanford University, Stanford, CA, 94305, USA.}
\address[label5]{California Institute for Quantitative Biosciences, UC Berkeley, Berkeley, CA 94720, USA}

\end{frontmatter}

\begin{figure*}[h]
 \centering
 \includegraphics[width=0.75\textwidth]{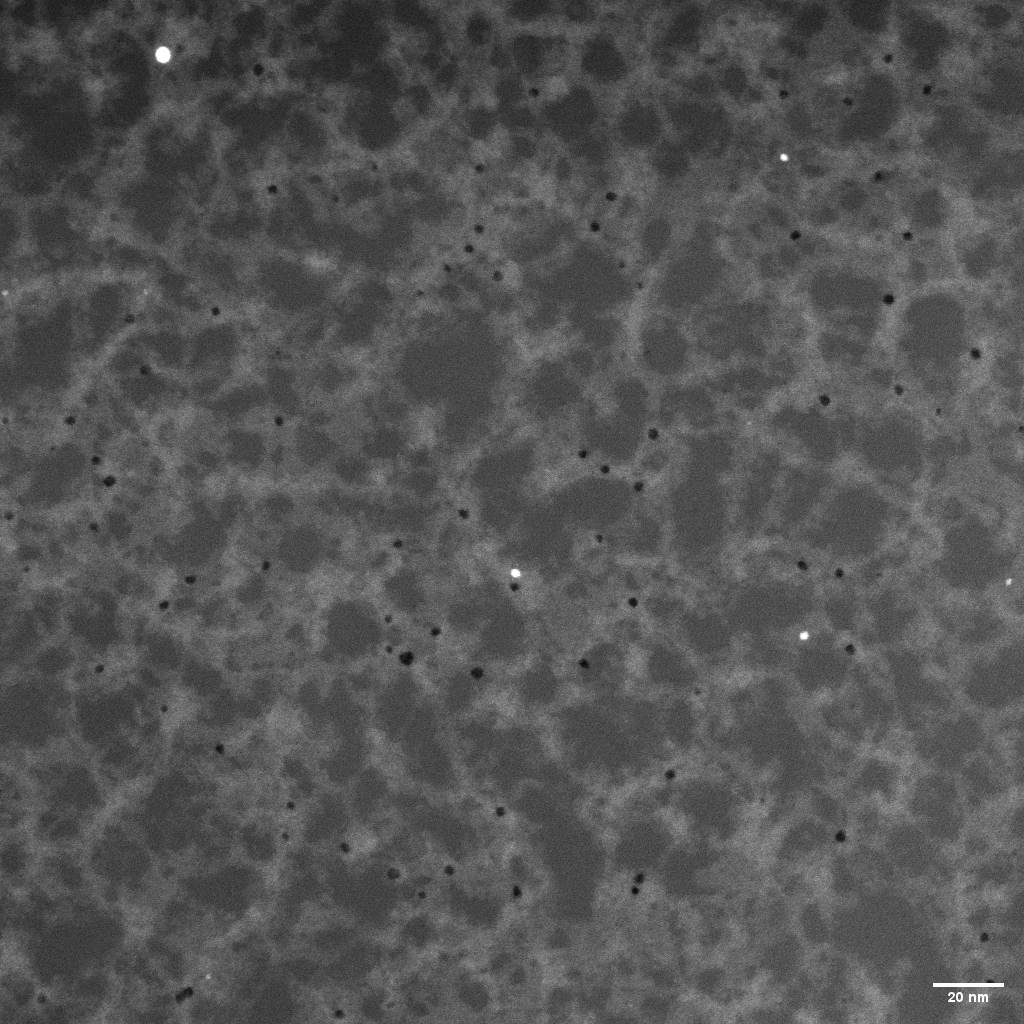}
 \caption{Low magnification HAADF-STEM image of $\sim$\qty{7}{nm} thick multilayer hBN that was exposed to the beam-blanked Ga LMIS for 20 minutes.}
 \label{Fig:low mag hBN STEM}
\end{figure*}

\begin{figure*}
 \centering
 \includegraphics[width=0.95\textwidth]{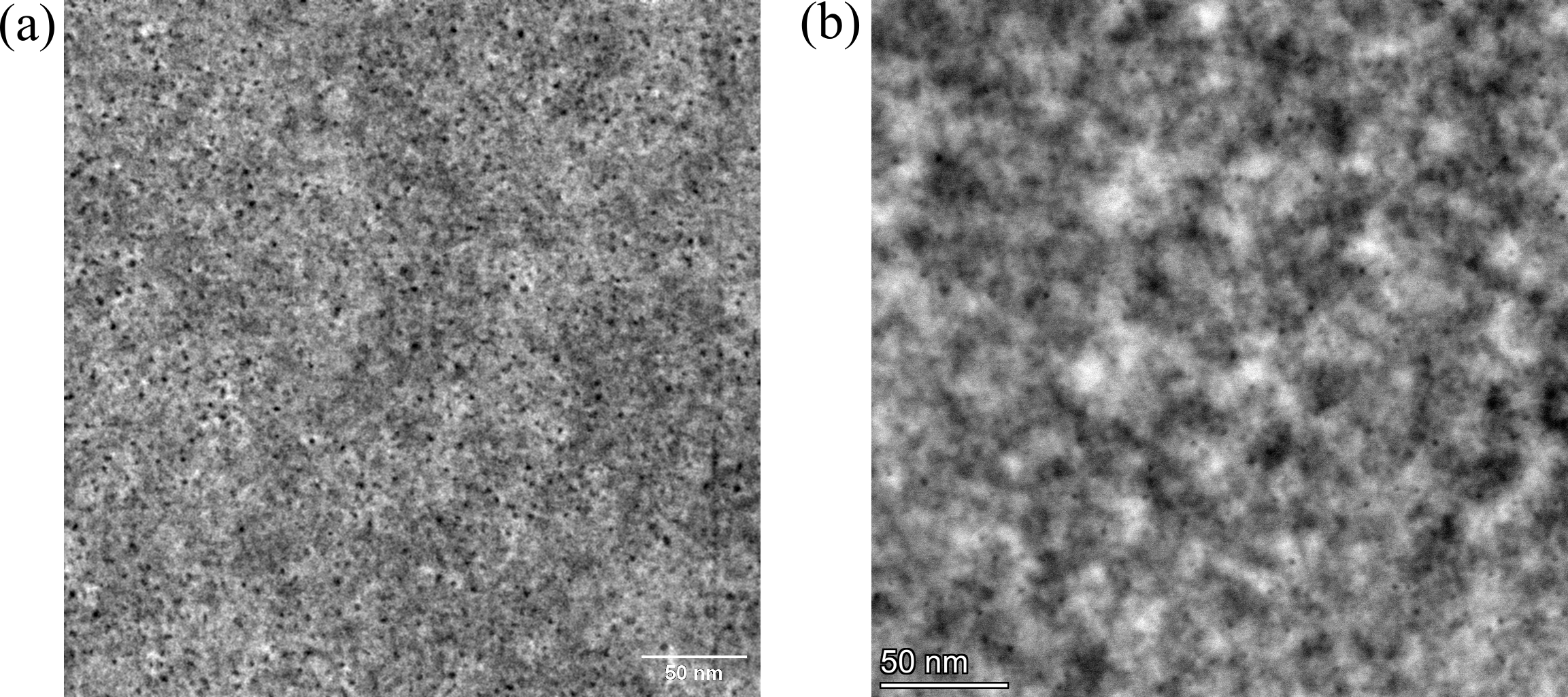}
 \caption{Low magnification HAADF-STEM images of (a) \qty{5}{nm} thick Si exposed to the beam-blanked Ga LMIS for 4 hours, and (b) \qty{20}{nm} thick SiN$_x$ exposed to the beam-blanked Ga LMIS for 40 minutes. The surface roughness of these samples gives rise to the more mottled contrast.}
 \label{Fig:low mag Si STEM}
\end{figure*}

 \begin{figure*} [hbt]
 \centering
 \includegraphics[width=0.85\textwidth]{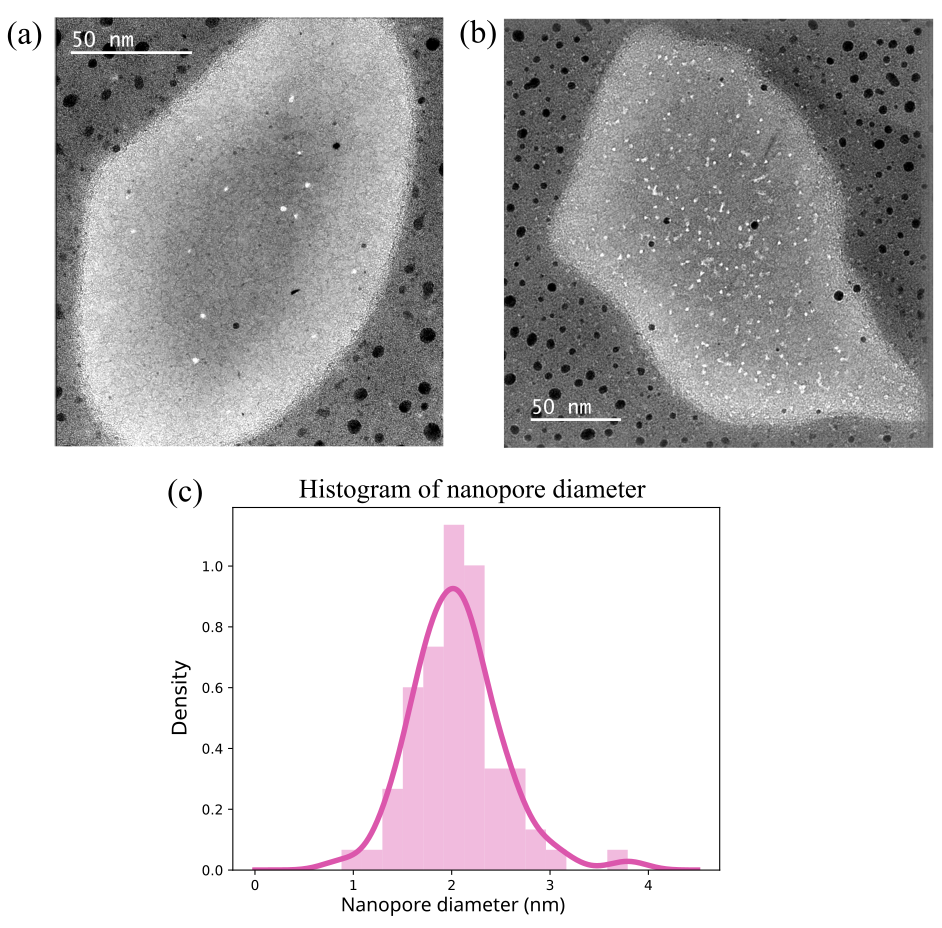}
 \caption{Bright-field TEM images of multilayer hBN (suspended over FIB-milled apertures in an Au-Pd sputter-coated SiN$_x$ membrane) exposed to the beam-blanked Ga LMIS for a newly replaced source for (a) 1 hour and (b) 20 hours. (c) Nanopore diameter histogram constructed from 75 nanopores found over several aperture regions of the sample pictured in (a). The curve represents the kernel density estimate plot.}
 \label{Fig:low mag Si STEM}
\end{figure*}

\begin{figure*} [hbt]
 \centering
 \includegraphics[width=0.95\textwidth]{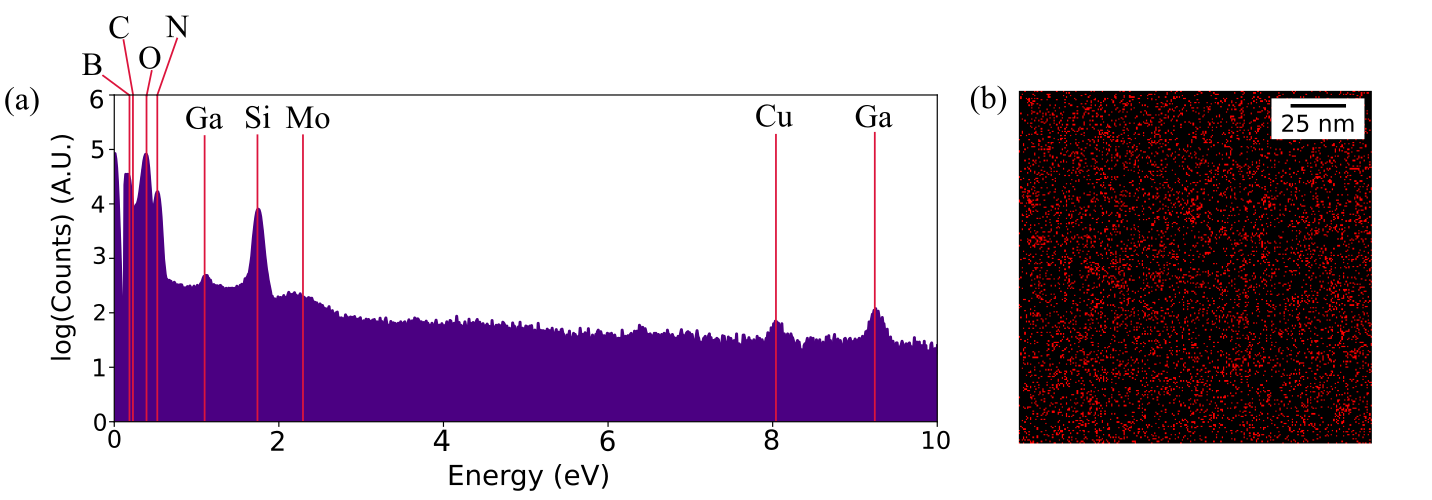}
 \caption{Further analysis of STEM-EDS results from Fig.~2 of the main manuscript. (a) Integrated EDS signal corresponding to the large field of view on the Ga-cluster-exposed hBN specimen. Below \qty{1}{keV} one finds the X-ray peaks for B and N, and also for C and O, which are expected as a result of surface contamination. The Si signal is likely a result of the SiN$_x$ support membrane, the Si/SiO$_2$ frame, and Si atoms from surface contamination. Cu and Mo trace signals are from scattering from the sample holder and microscope column components. (b) Corresponding elemental map for Ga, which due to the very low signal-to-noise ratio per pixel must be interpreted with caution.}
 \label{Fig:Si_SiN}
\end{figure*}
 
\begin{figure*} [hbt]
 \centering
 \includegraphics[width=0.8\textwidth]{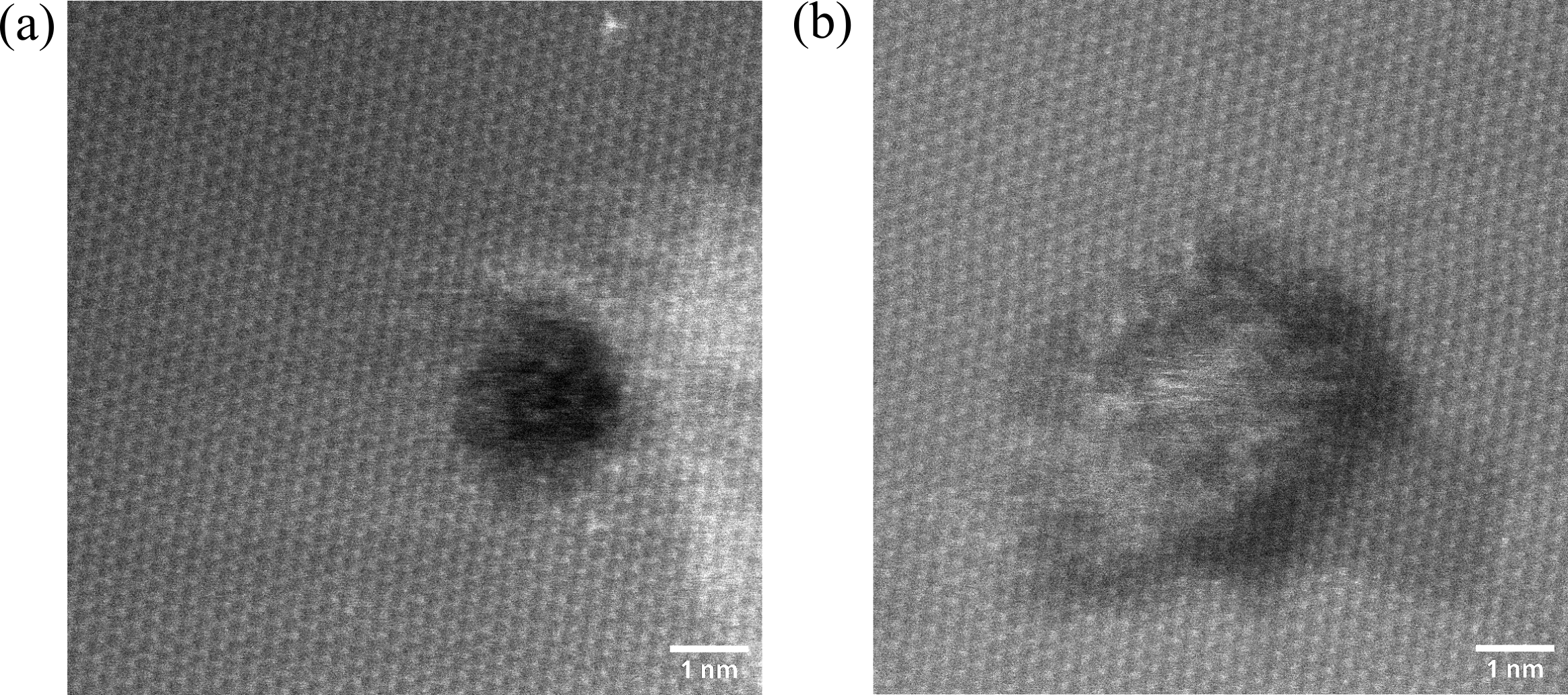}
 \caption{HAADF-STEM images of the (a) empty and (b) filled nanopores investigated by multislice ptychography in Fig.~4 of the main manuscript.}
 \label{Fig:Si_SiN}
\end{figure*}

\begin{figure*} [hbt]
 \centering
 \includegraphics[width=0.5\textwidth]{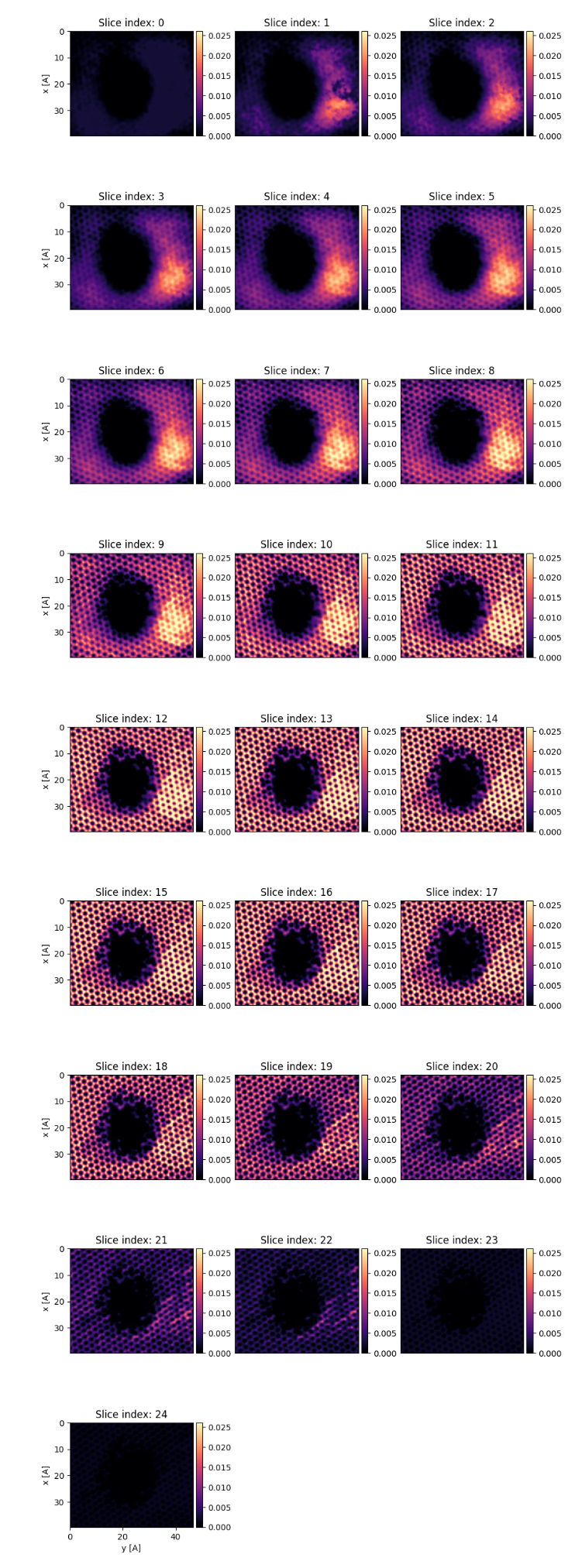}
 \caption{Full multislice ptychography reconstructions for the empty nanopore from Fig.~4 of the main manuscript.}
 \label{Fig:Si_SiN}
\end{figure*}

\begin{figure*} [hbt]
 \centering
 \includegraphics[width=0.5\textwidth]{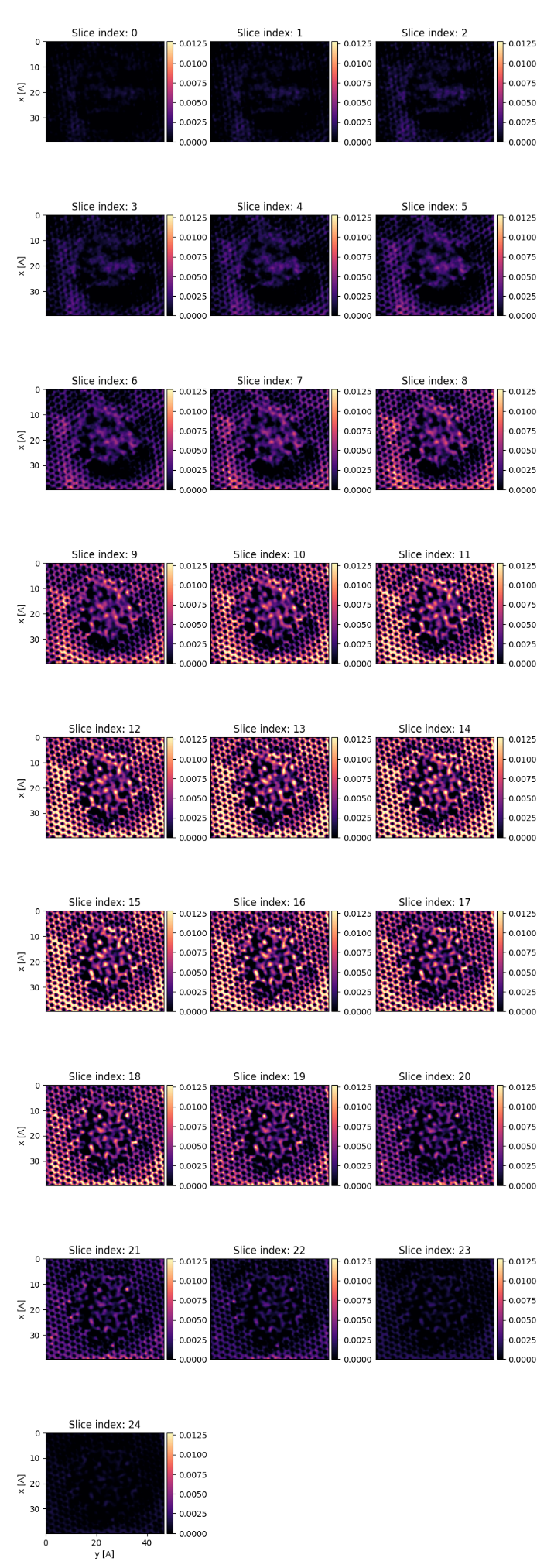}
 \caption{Full multislice ptychography reconstructions for the filled nanopore from Fig.~4 of the main manuscript.}
 \label{Fig:Si_SiN}
\end{figure*}

\begin{figure*} [hbt]
 \centering
 \includegraphics[width=0.7\textwidth]{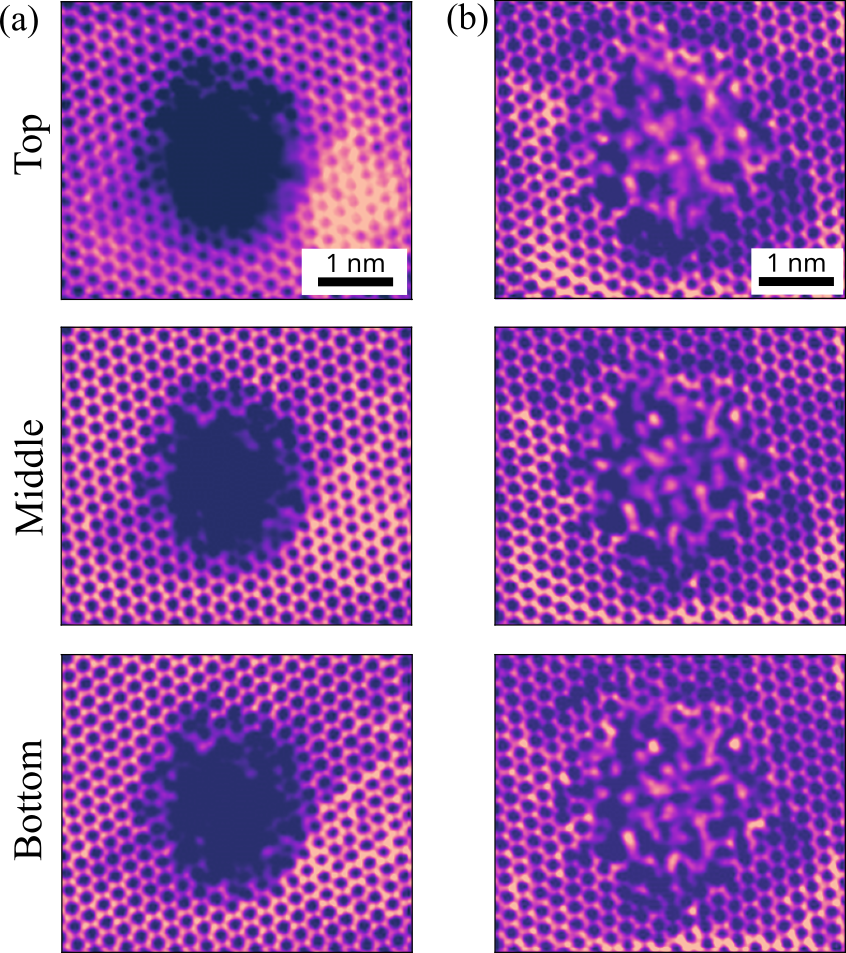}
 \caption{Reconstructed multislice ptychography slices corresponding to the top, middle, and bottom planes of the (a) empty and (b) filled nanopores from Fig.~4 of the main manuscript.}
 \label{Fig:Si_SiN}
\end{figure*}

\begin{figure*} [hbt]
 \centering
 \includegraphics[width=0.8\textwidth]{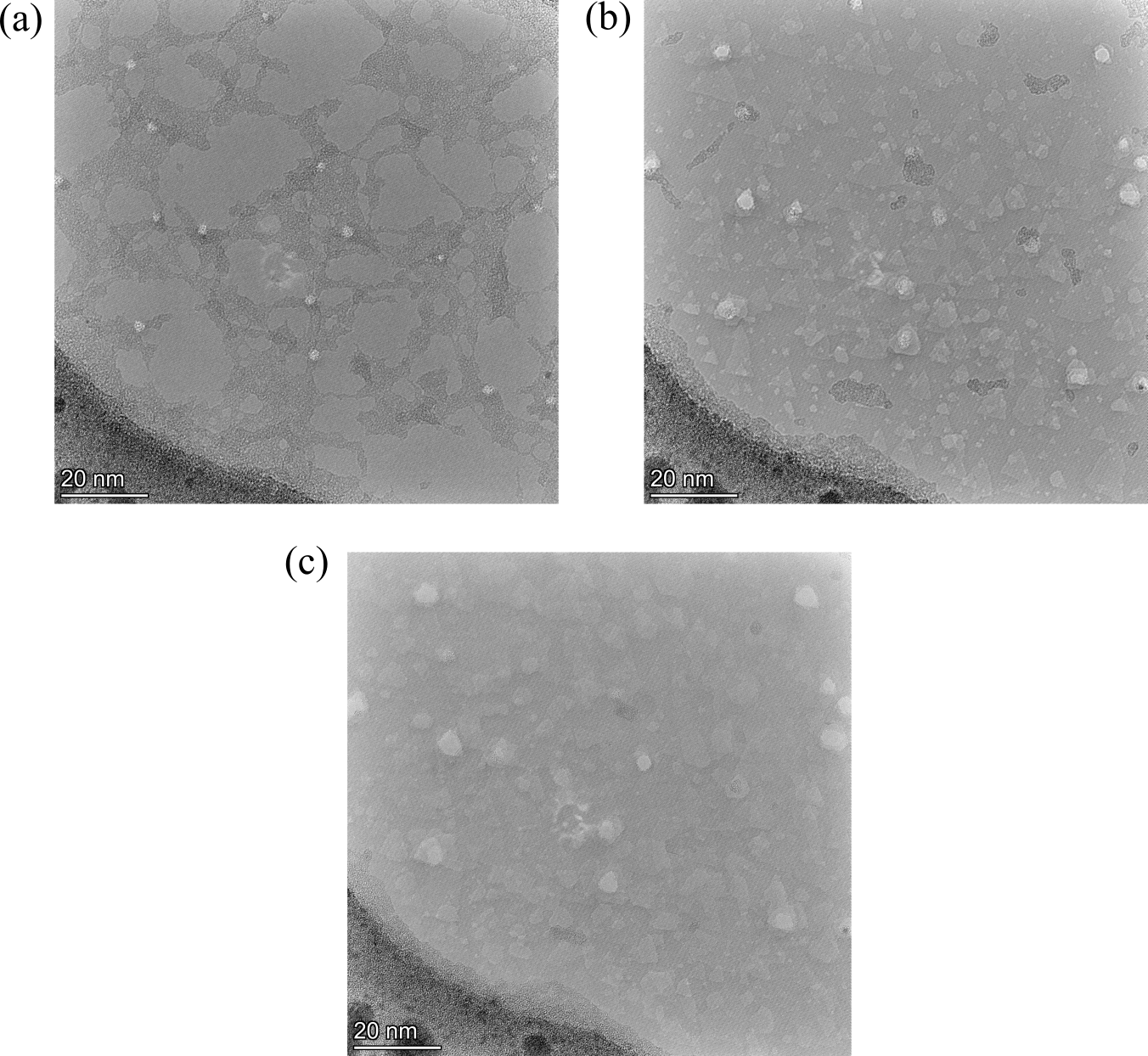}
 \caption{Full field of view HR-TEM images of nanopores from the unclogging and pore growth experiment summarized in Fig.~5 of the main manuscript. (a), (b), and (c) correspond to the $t = \qty{0}{s}, \qty{150}{s}, \qty{300}{s}$ time points.}
 \label{Fig:Si_SiN}
\end{figure*}
